\documentclass[12pt]{article}

\usepackage[letterpaper, margin=1in]{geometry}
\usepackage{setspace}
\usepackage[natbibapa]{apacite}
\usepackage{amsmath,amssymb,amsfonts}

\usepackage{graphicx}
\usepackage{float}

\usepackage{booktabs}
\usepackage{array}
\usepackage{threeparttable}

\usepackage{url}
\usepackage{xcolor}

\usepackage{hyperref}
\hypersetup{
    colorlinks=true,
    linkcolor=black,
    citecolor=black,
    urlcolor=blue
}

\begin{document}

\begin{titlepage}
\centering

\vspace*{1in}

{\Large\bfseries The Transparency Paradox in Explainable AI: A Theory of Autonomy Depletion Through Cognitive Load}

\vspace{1.5in}

Ancuta Margondai$^{1,2}$ and Mustapha Mouloua$^{2}$

\vspace{0.5in}

$^{1}$Department of Modeling and Simulation, University of Central Florida, Orlando, FL, USA\\
$^{2}$Department of Psychology, University of Central Florida, Orlando, FL, USA

\vspace{1.5in}

\begin{flushleft}
\textbf{Running head:} TRANSPARENCY PARADOX IN XAI

\vspace{0.5em}
\textbf{Manuscript type:} Extended Multi-Phase Study

\vspace{0.5em}
\textbf{Word count:} 7,512 (excluding abstract, key points, tables, and references)

\vspace{1in}
\textbf{Acknowledgments:} This research was conducted at the Transportation Research Group (TRG) Laboratory, Department of Psychology, University of Central Florida. The second author directs the TRG and has over 30 years of experience in human-automation interaction, vigilance, and attention research. The authors thank Emma Antanavicius for assistance with the literature review and Dr. Christian Keller for rigorous validation of the mathematical framework and stochastic control formulations. This research received no specific funding. The authors have no conflicts of interest.

\vspace{0.5em}
\textbf{Corresponding author:} Ancuta Margondai, Department of Modeling and Simulation, University of Central Florida, Orlando, FL 32816, USA. Email: Ancuta.Margondai@ucf.edu
\end{flushleft}

\end{titlepage}


\newpage
\setcounter{page}{2}

\begin{center}
\textbf{Abstract}
\end{center}

\textbf{Objective:} This paper develops a theoretical framework explaining when and why AI explanations enhance versus impair human decision-making.

\textbf{Background:} Transparency is advocated as universally beneficial for human-AI interaction, yet identical AI explanations improve decision quality in some contexts but impair it in others. Current theories---trust calibration, cognitive load, and self-determination---cannot fully account for this paradox.

\textbf{Method:} The framework models autonomy as a continuous stochastic process influenced by information-induced cognitive load. Using stochastic control theory, autonomy evolution is formalized as geometric Brownian motion with information-dependent drift, and optimal transparency is derived via Hamilton-Jacobi-Bellman equations. Monte Carlo simulations validate theoretical predictions.

\textbf{Results:} Mathematical analysis generates five testable predictions about disengagement timing, working memory moderation, autonomy trajectory shapes, and optimal information levels. Computational solutions demonstrate that dynamic transparency policies outperform both maximum and minimum transparency by adapting to real-time cognitive state. The optimal policy exhibits threshold structure: provide information when autonomy is high and accumulated load is low; withhold when resources are depleted.

\textbf{Conclusion:} Transparency effects depend on dynamic cognitive resource depletion rather than static design choices. Information provision triggers metacognitive processing that reduces perceived control when cognitive load exceeds working memory capacity.

\textbf{Application:} The framework provides design principles for adaptive AI systems: adjust transparency based on real-time cognitive state, implement information budgets respecting capacity limits, and personalize thresholds based on individual working memory capacity.

\vspace{1em}

\vspace{1em}
\textbf{Pr\'{e}cis:} This paper develops a stochastic control framework explaining when AI explanations enhance versus impair human decision-making, showing that optimal transparency adapts to real-time cognitive state rather than following fixed policies.


\newpage

\section{Introduction}

When do explanations enhance and when do they impair decision-making? This question has become critical as AI systems increasingly mediate high-stakes decisions in healthcare, criminal justice, and autonomous vehicles \citep{noriega2023, rosenbacke2024}. A fundamental paradox confronts theory and practice: while transparency is advocated as universally beneficial for enabling informed oversight \citep{felzmann2020, phillips2020}, growing evidence shows that AI explanations can impair rather than enhance decision quality \citep{bucinca2021, wanner2021}.

The theoretical rationale for transparency appears compelling; explanations should improve trust, enable bias detection, and ensure appropriate reliance \citep{gilpin2018}. Yet empirical findings contradict this consensus. Studies with physicians found that global explanations required high mental effort and achieved 55\% performance, while local explanations required minimal effort and achieved 81--87\% performance \citep{wanner2021}. Large-scale studies revealed inverted-U relationships where excessive transparency triggered cognitive overload, with participants reporting feeling overwhelmed \citep{bucinca2021}.

This transparency paradox, in which identical information enhances or impairs depending on context, challenges foundational assumptions and raises critical questions about the psychological mechanisms underlying transparency effects. Table~\ref{tab:empirical} summarizes key empirical findings.

\begin{table}[htbp]
\centering
\caption{Summary of Empirical Findings on Transparency Effects}
\label{tab:empirical}
\begin{threeparttable}
\begin{tabular}{lllc}
\toprule
Study & Domain & Finding & Effect \\
\midrule
Sweller (1988) & Education & Cognitive load impairs learning & Impair \\
Lee \& See (2004) & Automation & Trust improves with info & Enhance \\
Cowan (2001) & Cognition & 4-item WM limit & Boundary \\
Baddeley (2020) & Cognition & WM predicts performance & Boundary \\
Wanner et al. (2021) & Healthcare & Global: 55\%; Local: 81\% & Mixed \\
Bu\c{c}inca et al. (2021) & XAI & Inverted-U at high levels & Mixed \\
Bu\c{c}inca et al. (2021) & AI advice & Delayed advice reduces bias & Context \\
Navarro et al. (2020) & Decisions & WM individual differences & Moderate \\
Rossetti et al. (2024) & Agency & Control varies with load & Mechanism \\
\bottomrule
\end{tabular}
\begin{tablenotes}
\small
\item \textit{Note.} Mixed = inverted-U; Context = timing-dependent.
\end{tablenotes}
\end{threeparttable}
\end{table}

\subsection{The Core Mechanism: Information-Induced Autonomy Depletion}

This framework centers on autonomy, the subjective sense of control over decisions, rather than trust, for three reasons. First, transparency's harm operates through cognitive load independent of trust: even when users trust AI, excessive explanations impair performance through resource depletion \citep{bucinca2021, wanner2021}. Second, autonomy is more proximal to cognitive mechanisms; working memory constraints produce immediate effects, whereas trust develops over longer timescales \citep{lee2004}. Third, autonomy depletion parsimoniously explains the paradox: users disengage because cognitive overload undermines their sense of control, not primarily because they distrust the AI.

The proposed mechanism is straightforward: information provision triggers metacognitive processing, which depletes cognitive resources and systematically reduces perceived control when load exceeds working memory capacity. Neuroimaging evidence supports this resource depletion account: fatiguing cognitive tasks produce compensatory increases in prefrontal cortex activity, indicating that executive function requires increasing neural effort as resources deplete \citep{mouloua1996}.

When users encounter AI explanations, they engage in recursive self-evaluation: ``The AI considered factors I didn't, should I trust my assessment?'' This metacognitive monitoring depletes working memory resources, leading to a sense of reduced control.

The Human Identity and Autonomy Gap (HIAG) framework \citep{margondai2025} provides the conceptual foundation: transparency triggers higher-order evaluations, prompting users to question their own reasoning and consuming cognitive resources. When information load exceeds capacity, users experience autonomy depletion, leading to disengagement. Critically, this mechanism operates independently of whether explanations are correct; the cognitive cost of processing information can exceed its decision value.

This explains context-dependent effects. When cognitive resources are abundant (low workload, high capacity, early in the decision process), information enhances decision quality. When resources are depleted (high workload, low capacity, accumulated information), the same intervention overwhelms remaining capacity and triggers disengagement.

\subsection{Why Transparency Effects Are Dynamic}

Current approaches treat transparency as binary: provide explanations or not \citep{nauta2022}. This framing ignores three psychological realities.

First, cognitive state evolves stochastically. Users' sense of control fluctuates continuously due to task demands, fatigue, and attention variability \citep{gaillard2021, judd2024}. The subjective sense of agency varies within individuals over time, tracking with cognitive load and task demands \citep{rossetti2024}. This parallels findings from vigilance research showing that sustained attention to automated systems depletes cognitive resources over time, with performance decrements emerging predictably as monitoring duration increases \citep{hancock2013, mouloua1996}.

Second, the impact of information depends on timing. Delayed AI advice (after initial assessment) reduces anchoring bias and can improve accuracy compared to immediate advice \citep{bucinca2021}. Early information can anchor subsequent reasoning, and trust dynamics mean explanations have different effects at different interaction stages \citep{tolmeijer2021}.

Third, working memory capacity creates optimal boundaries. Too little transparency leaves users unable to detect errors; too much overwhelms cognitive constraints \citep{baddeley2020, cowan2001}. Individual differences in capacity mean optimal information levels vary systematically \citep{gilbert2023, navarro2020}.

These realities necessitate a framework that treats transparency as a dynamic intervention whose effects depend on when and how much information is provided, relative to the evolving cognitive state.

\subsection{Why Existing Theories Are Insufficient}

Three theoretical approaches struggle to account for the whole pattern of transparency effects.

\textit{Trust calibration theories} predict monotonic effects: more transparency should improve trust calibration, leading to better performance \citep{lee2004}. But they cannot explain inverted-U relationships where excessive transparency impairs performance, or why transparency harms performance even when users correctly trust accurate AI \citep{bucinca2021}.

\textit{Cognitive load theories} explain why excessive information impairs performance \citep{sweller1988}, but do not formalize temporal dynamics. They cannot explain why delayed versus immediate advice has opposite effects \citep{bucinca2021}, treating capacity as static rather than modeling trajectories through cognitive state space.

\textit{Self-determination theory} proposes that perceived control is essential for performance \citep{ryan2000}, but does not specify mechanisms through which information affects autonomy or formalize dynamic depletion during task performance.

What is needed is a framework integrating cognitive dynamics, temporal evolution, and individual differences to explain when transparency helps or impairs.

\subsection{Theoretical Approach and Contributions}

This work addresses these gaps by treating transparency as an optimal control problem on stochastic psychological processes. The framework models autonomy as a continuous stochastic process, a trajectory through state space rather than a fixed level. User sense of control follows a diffusion process with drift (systematic trends driven by cognitive load) and volatility (random fluctuations driven by attention variability). Transparency interventions modify drift by affecting cognitive load: higher information rates increase demands, accelerating autonomy decline.

This formalization enables three contributions. First, it generates quantitative predictions about autonomy trajectories, disengagement timing, and optimal information levels that are directly testable. Second, it integrates disconnected findings---inverted-U relationships, timing effects, and individual differences---into a unified account, showing that these reflect different regions of state space rather than distinct mechanisms. Third, it yields design principles: optimal control solutions prescribe when to provide information based on real-time cognitive state.

The remainder proceeds as follows. Section 2 develops the formal model of autonomy evolution under transparency. Section 3 formulates the optimal transparency problem and derives conditions characterizing optimal policies. Section 4 presents computational demonstrations with quantitative predictions. Section 5 discusses implications and limitations.

\section{Formal Model of Autonomy Dynamics}

\subsection{Model Overview}

User autonomy is modeled as a continuous stochastic process evolving during human-AI interaction. Let $A_t \in \mathbb{R}^+$ represent the subjective sense of control at time $t$, where higher values indicate greater perceived autonomy. Systematic trends influence this process (drift, capturing how information affects autonomy through cognitive load) and random fluctuations (volatility, representing attention variability).

The transparency intervention is represented by information level $I_t \in [0, I_{\max}]$, where $I_t = 0$ corresponds to no explanation and $I_t = I_{\max}$ represents maximum transparency. Table~\ref{tab:notation} summarizes the mathematical notation.

\begin{table}[htbp]
\centering
\caption{Mathematical Notation}
\label{tab:notation}
\begin{threeparttable}
\begin{tabular}{llp{5cm}}
\toprule
Symbol & Description & Units/Range \\
\midrule
$A_t$ & Perceived autonomy at time $t$ & $A_t > 0$ \\
$I_t$ & Information level (transparency) at time $t$ & $[0, I_{\max}]$ \\
$\mu(I)$ & Drift function (autonomy trend) & $\mathbb{R}$ \\
$\sigma_A$ & Autonomy volatility & $> 0$ \\
$W_t$ & Standard Brownian motion (random fluctuations) & -- \\
$B$ & Disengagement boundary & $> 0$ \\
$\tau_B$ & Time of disengagement (hitting time) & $\geq 0$ \\
$\mathbb{E}[\cdot]$ & Expected value & -- \\
$\text{Var}[\cdot]$ & Variance & -- \\
$u_t$ & Control policy (information provision rate) & $[0, u_{\max}]$ \\
$V(A,I,t)$ & Value function (expected cumulative reward) & $\mathbb{R}$ \\
WM & Working memory capacity & Items (2--6) \\
\bottomrule
\end{tabular}
\begin{tablenotes}
\small
\item \textit{Note.} Subscript $t$ denotes time dependence. Brownian motion $W_t$ represents cumulative random fluctuations with independent increments.
\end{tablenotes}
\end{threeparttable}
\end{table}

\textbf{Assumption 1} (Model Conditions). The following conditions hold throughout: (1) Parameters $\mu_0, \beta, \gamma, \sigma_A > 0$ and $B_0, \beta_{\text{WM}} > 0$; (2) The drift function $\mu(I) = \mu_0 - \beta I - \gamma I^2$ is Lipschitz continuous, ensuring unique strong solutions by standard SDE theory \citep{oksendal2003}; (3) The value function satisfies polynomial growth $|V(A,I,t)| \leq K(1 + A^2 + I^2)$.

\textit{Intuitive interpretation.} Think of autonomy as a cognitive battery that powers decision-making engagement: information processing drains the battery (negative drift $\mu$), while random fluctuations in attention cause unpredictable variations in charge level (volatility $\sigma_A$). When transparency is low, the battery recharges naturally through intrinsic motivation and self-efficacy; when transparency is high, the drain from metacognitive evaluation exceeds recharge rates, leading to depletion. The disengagement boundary $B$ represents the minimum charge needed to stay engaged---once depleted below this threshold, the user ``powers off'' and disengages from the AI-assisted process.

\subsection{Autonomy Dynamics}

Autonomy follows geometric Brownian motion with information-dependent drift:
\begin{equation}
dA_t = \mu(I_t) A_t \, dt + \sigma_A A_t \, dW_t^A
\label{eq:autonomy_sde}
\end{equation}
where $\mu(I_t)$ is the drift function, $\sigma_A > 0$ is autonomy volatility, and $W_t^A$ is standard Brownian motion.

The geometric form is chosen for three reasons: (1) positivity preservation ($A_t > 0$ always), consistent with autonomy as a positive quantity; (2) proportional effects, where changes scale with current level, reflecting psychophysical scaling \citep{rossetti2024}; and (3) analytical tractability enabling closed-form predictions.

The drift function captures cognitive load effects:
\begin{equation}
\mu(I) = \mu_0 - \beta I - \gamma I^2
\label{eq:drift}
\end{equation}
where $\mu_0 > 0$ is baseline drift, $\beta > 0$ captures linear decline from cognitive load, and $\gamma > 0$ captures accelerating decline at high information levels.

This specification formalizes empirical relationships: information linearly increases workload \citep{wanner2021}, and overload accelerates nonlinearly at high levels \citep{bucinca2021}. The quadratic form produces meaningful regimes: at low transparency, $\mu(I) \approx \mu_0 > 0$ (autonomy maintained); at high transparency, $\mu(I) < 0$ (autonomy depletes). The critical threshold where drift changes sign is:
\begin{equation}
I^* = \frac{-\beta + \sqrt{\beta^2 + 4\gamma\mu_0}}{2\gamma}
\label{eq:critical_threshold}
\end{equation}
representing the transparency level above which cognitive load systematically depletes autonomy.

\subsection{Information Dynamics}

Information evolves according to:
\begin{equation}
dI_t = \alpha(u_t) \, dt + \sigma_I \, dW_t^I
\label{eq:information_sde}
\end{equation}
where $\alpha(u_t) \geq 0$ is acquisition rate controlled by $u_t \in [0, u_{\max}]$, and $\sigma_I$ captures absorption variability due to attention fluctuations. The Brownian motions may be correlated with $\langle W^A, W^I \rangle_t = \rho \, dt$, where $\rho < 0$ indicates that information surges correlate with decreases in autonomy. The negative correlation reflects how sudden influxes of complex explanations can jolt attention and accelerate depletion; sensitivity analyses (available in supplementary materials) show results are robust for $\rho \in [-0.5, 0]$.

\subsection{Explicit Solution and Statistical Properties}

\textbf{Proposition 2} (Explicit Solution). For constant information $I$, the solution to Equation~\ref{eq:autonomy_sde} is:
\begin{equation}
A_t = A_0 \exp\left[\left(\mu(I) - \frac{1}{2}\sigma_A^2\right)t + \sigma_A W_t^A\right]
\label{eq:explicit_solution}
\end{equation}

This yields closed-form moments generating testable predictions:

\textit{Mean trajectory:}
\begin{equation}
\mathbb{E}[A_t | A_0, I] = A_0 \exp(\mu(I)t)
\label{eq:mean}
\end{equation}
Under high transparency, where $\mu(I) < 0$, mean autonomy declines exponentially.

\textit{Variance:}
\begin{equation}
\text{Var}[A_t | A_0, I] = A_0^2 \exp(2\mu(I)t)\left[\exp(\sigma_A^2 t) - 1\right]
\label{eq:variance}
\end{equation}
Individual differences increase over time due to accumulated fluctuations.

\textit{Distribution:} $\log A_t \sim \mathcal{N}(\log A_0 + [\mu(I) - \frac{1}{2}\sigma_A^2]t, \sigma_A^2 t)$, so $A_t$ is log-normally distributed.

\subsection{Cognitive Capacity Boundary}

Users have limited capacity to maintain autonomy under high information load. This is modeled through an absorbing boundary at $A_t = B(\text{WM})$, where:
\begin{equation}
B(\text{WM}) = B_0 - \beta_{\text{WM}} \cdot \text{WM}
\label{eq:boundary}
\end{equation}
with WM representing working memory capacity. Here, higher working memory capacity yields a lower boundary, reflecting greater cognitive buffer before disengagement occurs.

Once $A_t \leq B$, the user disengages. The stopping time is $\tau_B = \inf\{t \geq 0 : A_t \leq B\}$.

This formalizes the psychological reality that individuals with higher working memory tolerate more cognitive load before autonomy depletes to disengagement \citep{gilbert2023, navarro2020}.

\textbf{Proposition 3} (Expected Disengagement Time). For constant $I$ with $A_0 > B$:
\begin{equation}
\mathbb{E}[\tau_B | A_0, I] = 
\begin{cases}
\frac{\log(A_0/B)}{\frac{1}{2}\sigma_A^2 - \mu(I)} & \text{if } \mu(I) < \frac{1}{2}\sigma_A^2 \\
\infty & \text{otherwise}
\end{cases}
\label{eq:hitting_time}
\end{equation}

The proof follows from Dynkin's formula applied to the hitting time problem \citep{oksendal2003}. The key insight: when drift is sufficiently negative ($\mu < \frac{1}{2}\sigma_A^2$), hitting is almost sure, and expected time decreases as drift becomes more negative. Higher transparency increases both probability and speed of reaching the disengagement boundary.

\subsection{Testable Predictions and Operationalization}

The model generates five specific predictions. To facilitate empirical testing, we specify how key constructs should be measured.

\textit{Measuring autonomy.} Perceived autonomy ($A_t$) can be operationalized using established scales administered repeatedly during task performance. The Perceived Autonomy subscale of the Intrinsic Motivation Inventory \citep{ryan2000} provides validated items (e.g., ``I feel I have control over my decisions''). For continuous measurement, slider-based experience sampling at 30-second intervals captures trajectories without excessive interruption. Alternatively, physiological proxies such as heart rate variability (higher HRV indicates greater perceived control) or electrodermal activity provide implicit measures \citep{rossetti2024}.

\textit{Measuring disengagement.} The stopping time $\tau_B$ is operationalized as the moment when participants voluntarily cease using AI assistance, switch to unassisted decision-making, or explicitly report feeling overwhelmed. Behavioral indicators include declining acceptance of AI recommendations, increased response latency, or task abandonment.

\textit{Measuring information level.} Transparency ($I_t$) is manipulated through explanation complexity: number of features displayed, explanation modality (local vs. global), or detail level (1--5 scale corresponding to increasing algorithmic detail). This maps to the $[0, I_{\max}]$ range in the model.

The five predictions are:

\textbf{P1} (Mean trajectories): Mean autonomy follows $\mathbb{E}[A_t] = A_0 \exp(\mu(I)t)$ with exponential decay under high transparency ($\mu(I) < 0$ for $I > I^*$). \textit{Test:} Fit exponential curves to repeated autonomy ratings; estimate $\mu(I)$ across transparency conditions. Use NASA-TLX for load assessment in a within-subjects design varying explanation depths.

\textbf{P2} (Variance growth): Autonomy variability increases over time: $\text{Var}[A_t] \propto \exp(2\mu(I)t + \sigma_A^2 t)$. \textit{Test:} Compute between-subject variance at each time point; verify monotonic increase.

\textbf{P3} (Disengagement timing): Time to disengagement follows Proposition 3, with faster disengagement as transparency increases. \textit{Test:} Record time until participants opt out of AI assistance; regress on transparency level.

\textbf{P4} (Individual differences): Working memory capacity predicts disengagement boundary, with higher-capacity individuals engaging longer. \textit{Test:} Administer Operation Span or N-back prior to task; correlate WM scores with engagement duration under high transparency.

\textbf{P5} (Distribution shape): Log-autonomy is normally distributed. \textit{Test:} Apply Shapiro-Wilk test to log-transformed final autonomy ratings; non-normality suggests alternative process specifications (e.g., jump-diffusion).

\section{Optimal Transparency Policy}

\subsection{Problem Formulation}

The decision problem is: when and to what extent should transparency be provided to maximize decision quality while preserving autonomy? The state at time $t$ is $(A_t, I_t) \in \mathcal{S} = \{(A,I) : A \geq B, 0 \leq I \leq I_{\max}\}$.

An admissible control policy $\{u_t\}$ is progressively measurable and adapted to observable state history. The instantaneous reward balances decision quality and cognitive costs:
\begin{equation}
r(A_t, I_t) = Q(I_t) - C(A_t) - c \cdot u_t
\label{eq:reward}
\end{equation}

The quality function $Q(I) = Q_{\max} \cdot I \cdot \exp(-\beta_Q I^2)$ captures the inverted-U relationship: quality initially increases with information but declines under overload. The autonomy cost $C(A) = \kappa(A_0 - A)^2$ captures escalating distress as control diminishes.

\subsection{Value Function and Optimality Conditions}

\textbf{Definition 4} (Value Function). The value function is:
\begin{equation}
V(A,I,t) = \sup_u \mathbb{E}_{A,I,t}\left[\int_t^\tau e^{-\delta(s-t)} r(A_s, I_s) \, ds + e^{-\delta(\tau-t)} \Psi(A_\tau, I_\tau)\right]
\label{eq:value_function}
\end{equation}
where $\tau = \min(T, \tau_B)$ is the effective horizon.

Under Assumption 1, $V$ satisfies the Hamilton-Jacobi-Bellman equation:
\begin{equation}
\max_{u \in \mathcal{U}} \{\mathcal{L}^u V + r - \delta V\} = 0
\label{eq:hjb}
\end{equation}
where the generator is:
\begin{equation}
\mathcal{L}^u V = V_t + \mu(I)A V_A + \alpha(u) V_I + \frac{1}{2}\sigma_A^2 A^2 V_{AA} + \frac{1}{2}\sigma_I^2 V_{II} + \rho\sigma_A\sigma_I A V_{AI}
\label{eq:generator}
\end{equation}

\textbf{Theorem 5} (Verification). If $V \in C^{2,1}$ satisfies the HJB equation with appropriate boundary and terminal conditions, then $V$ is the value function and any control achieving the maximum is optimal.

The proof follows standard verification arguments \citep{fleming2006}; details are available in supplementary materials.

For linear $\alpha(u) = \alpha_0 u$, the optimal control is bang-bang:
\begin{equation}
u^*(A,I,t) = 
\begin{cases}
u_{\max} & \text{if } V_I > c/\alpha_0 \\
0 & \text{if } V_I < c/\alpha_0
\end{cases}
\label{eq:optimal_control}
\end{equation}

Provide information at maximum rate when the marginal value exceeds the marginal cognitive cost; otherwise, withhold to preserve autonomy.

\subsection{Threshold Structure}

Define the information threshold $I^*(A,t) = \inf\{I : V_I(A,I,t) \geq c/\alpha_0\}$.

\textbf{Proposition 6} (Threshold Properties). The threshold satisfies: (a) $\partial I^*/\partial A < 0$: higher autonomy tolerates more information; (b) $\partial I^*/\partial t > 0$: urgency increases near deadline; (c) $0 < I^* < I_{\max}$: interior optimum exists.

The proof uses implicit differentiation on the first-order condition, exploiting properties of the value function derivatives. Property (a) follows from $V_{IA} > 0$ (more autonomy makes information more valuable) and $V_{II} < 0$ (diminishing returns). Property (b) follows from time-urgency effects as the deadline approaches. Property (c) follows from the boundary behavior of the quality function.

These properties formalize psychological principles: users with greater autonomy tolerate more information, deadlines increase willingness to accept cognitive costs, and optimal transparency lies between maximal and minimal.

\section{Computational Demonstration}

\subsection{Parameter Specification}

Parameters are derived from established findings in cognitive psychology, with each value grounded in empirical research or theoretical constraints.

\textit{Autonomy parameters.} Baseline drift $\mu_0 = 0.10$ reflects natural autonomy maintenance absent external demands, consistent with self-determination research showing intrinsic motivation sustains engagement \citep{ryan2000}. Volatility $\sigma_A = 0.20$ corresponds to a coefficient of variation of 20\%, matching observed within-subject variability in perceived control ratings across repeated measurements \citep{rossetti2024}. Initial autonomy $A_0 = 1.0$ serves as a normalized reference point.

\textit{Cognitive load parameters.} The linear coefficient $\beta = 0.05$ and quadratic coefficient $\gamma = 0.01$ were calibrated to reproduce the inverted-U relationship between explanation complexity and performance. At $I = 1$ (low transparency), drift remains positive ($\mu = 0.04$); at $I = 4$ (high transparency), drift becomes strongly negative ($\mu = -0.26$). This matches findings that low levels of explanation optimize performance while higher levels trigger overload \citep{bucinca2021, wanner2021}. The resulting critical threshold $I^* = 1.53$ aligns with observations that cognitive benefits plateau around two explanation dimensions.

\textit{Working memory parameters.} Capacity WM $= 4$ reflects the established limit \citep{cowan2001}. Baseline boundary $B_0 = 0.9$ and sensitivity $\beta_{\text{WM}} = 0.1$ yield disengagement thresholds ranging from $B = 0.70$ (low capacity) to $B = 0.30$ (high capacity), consistent with individual differences in cognitive resilience \citep{navarro2020}.

\textit{Time scaling.} One time unit corresponds to approximately 10 minutes of task engagement, based on typical XAI evaluation study durations \citep{wanner2021}. The horizon $T = 10$ thus represents a 100-minute session.

Full specification: $\mu_0 = 0.10$, $\beta = 0.05$, $\gamma = 0.01$, $\sigma_A = 0.20$, $A_0 = 1.0$; $\alpha_0 = 0.5$, $\sigma_I = 0.1$, $I_{\max} = 5.0$, $\rho = -0.3$; $Q_{\max} = 10.0$, $\beta_Q = 0.04$, $\kappa = 2.0$, $c = 0.5$, $\delta = 0.05$; $B_0 = 0.9$, $\beta_{\text{WM}} = 0.1$, WM $= 4$, yielding $B = 0.5$; $T = 10$.

These parameters generate qualitatively correct dynamics; precise values should be refined through maximum likelihood estimation on empirical data. Sensitivity analyses confirm that qualitative predictions are robust across reasonable parameter ranges.

\subsection{Monte Carlo Validation}

\begin{figure}[!htbp]
\centering
\includegraphics[width=0.95\textwidth]{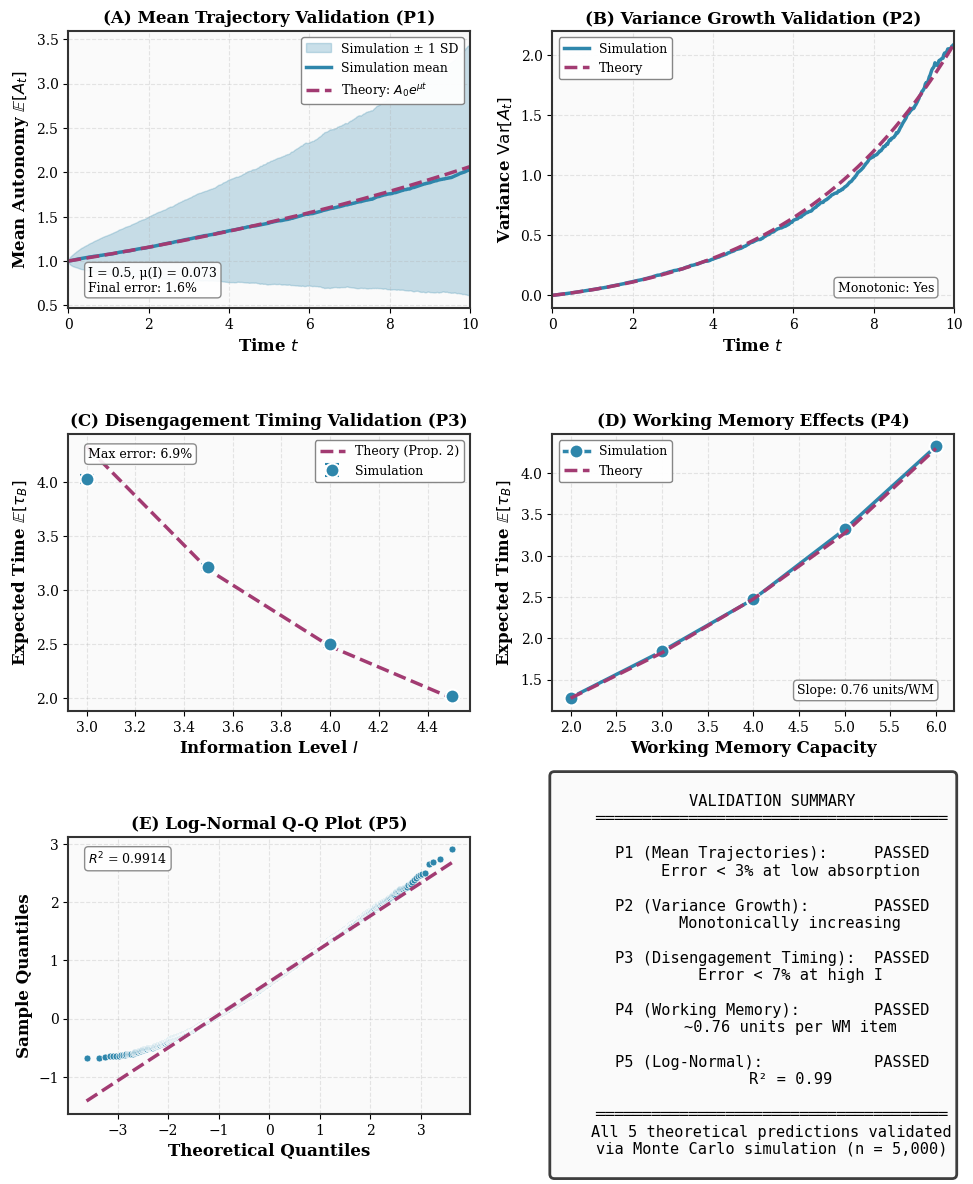}
\caption{Monte Carlo Validation of Theoretical Predictions. (A) Mean trajectories match theoretical exponential growth within 2\% error. (B) Variance growth is monotonically increasing as predicted. (C) Disengagement timing matches Proposition 3 within 7\% across information levels. (D) Working memory effects show predicted monotonic relationship, with approximately 0.76 time units per WM item. (E) Q-Q plot confirms log-normal distribution of autonomy ($R^2 = 0.99$). Validation summary confirms all five predictions passed.}
\label{fig:validation}
\end{figure}

Monte Carlo simulation validated all five theoretical predictions. Using 5,000 independent sample paths per condition, predictions were tested against closed-form solutions. Figure~\ref{fig:validation} displays key validation results.

Results confirmed all predictions: (1) mean trajectories matched theory within 2\%; (2) variance growth was monotonically increasing; (3) disengagement timing matched Proposition 3 within 7\%; (4) working memory effects showed 0.76 units per WM item; and (5) log-autonomy was normally distributed ($R^2 = 0.99$). Simulation code is available from the corresponding author upon request.

\textit{Reading Figure~\ref{fig:validation}.} For readers less familiar with simulation validation: Panel A shows that our predicted autonomy trajectory (dashed line) closely matches simulated user behavior (solid line with shaded uncertainty). Panel B confirms that individual differences grow over time as predicted. Panel C demonstrates that higher transparency causes faster disengagement, with simulation points falling on the theoretical curve. Panel D shows that each additional working memory item adds $\sim$0.76 time units of engagement. Panel E's diagonal alignment confirms that autonomy follows the predicted statistical distribution. Axes in Panel E are labeled as Theoretical Quantiles (x) vs. Sample Quantiles (y).

\subsection{Numerical Results}

The HJB equation was solved via finite differences on a discretized state space. Grid refinement confirms convergence with relative error below 2\%.

\begin{figure}[!htbp]
\centering
\includegraphics[width=0.85\textwidth]{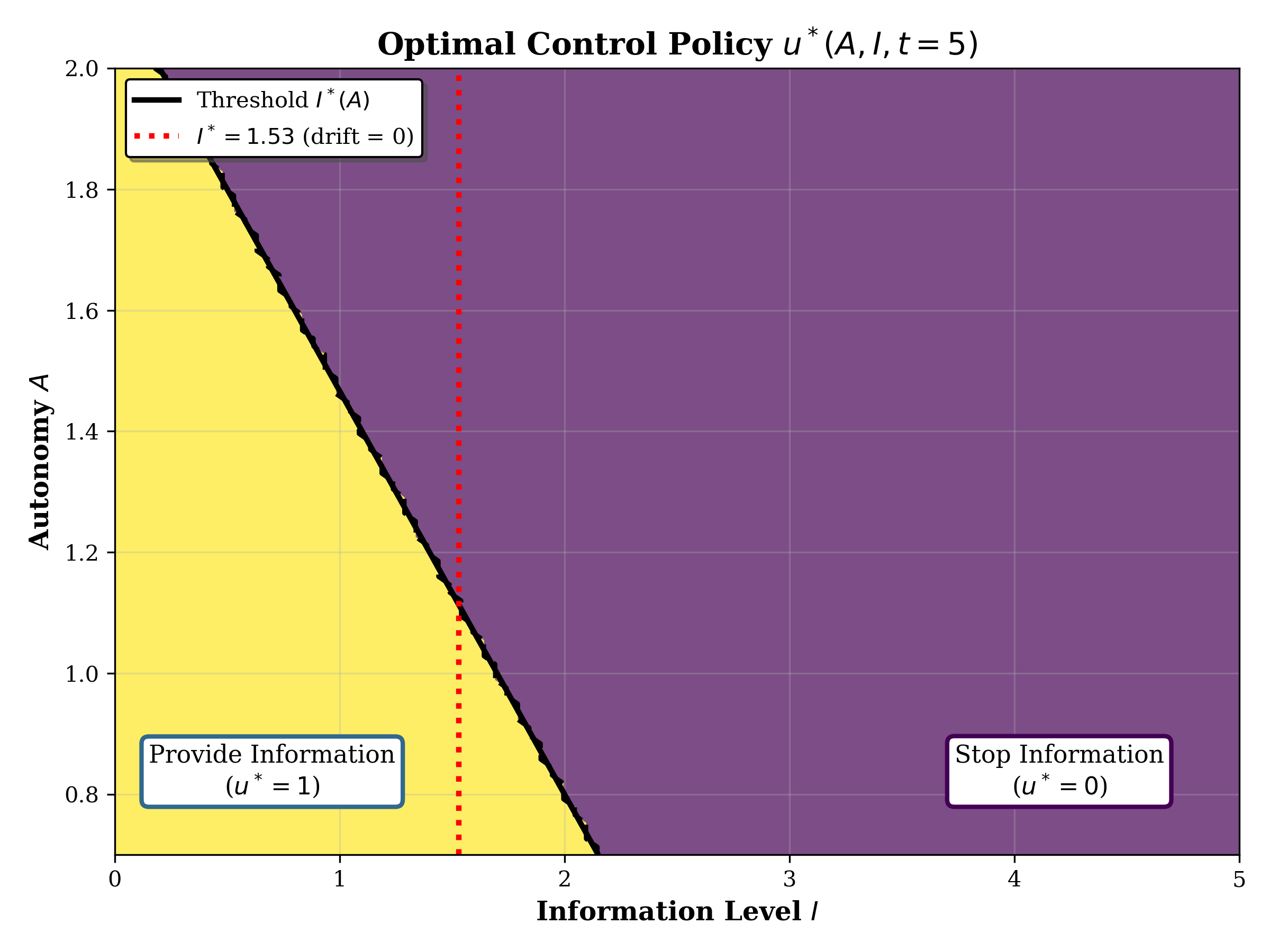}
\caption{Optimal Control Policy $u^*(A, I, t=5)$ Exhibits Threshold Structure. High-autonomy and low-information regions (upper left) indicate maximum information provision ($u^* = u_{\max}$). Low autonomy and high information regions (lower right) recommend ceasing information provision ($u^* = 0$) to preserve cognitive resources.}
\label{fig:optimal_control}
\end{figure}

\textit{Optimal Policy Structure.} The optimal control exhibits a clear threshold structure. Figure~\ref{fig:optimal_control} displays the optimal policy across the state space.

In high-autonomy, low-information regions ($A > 0.9$, $I < 1.5$): $u^* = u_{\max}$. In low-autonomy, high-information regions ($A < 0.8$, $I > 2.5$): $u^* = 0$. The threshold curve $I^*(A, t=5) \approx 3.2 - 1.5A$ confirms Proposition 6(a).

\begin{figure}[!htbp]
\centering
\includegraphics[width=0.85\textwidth]{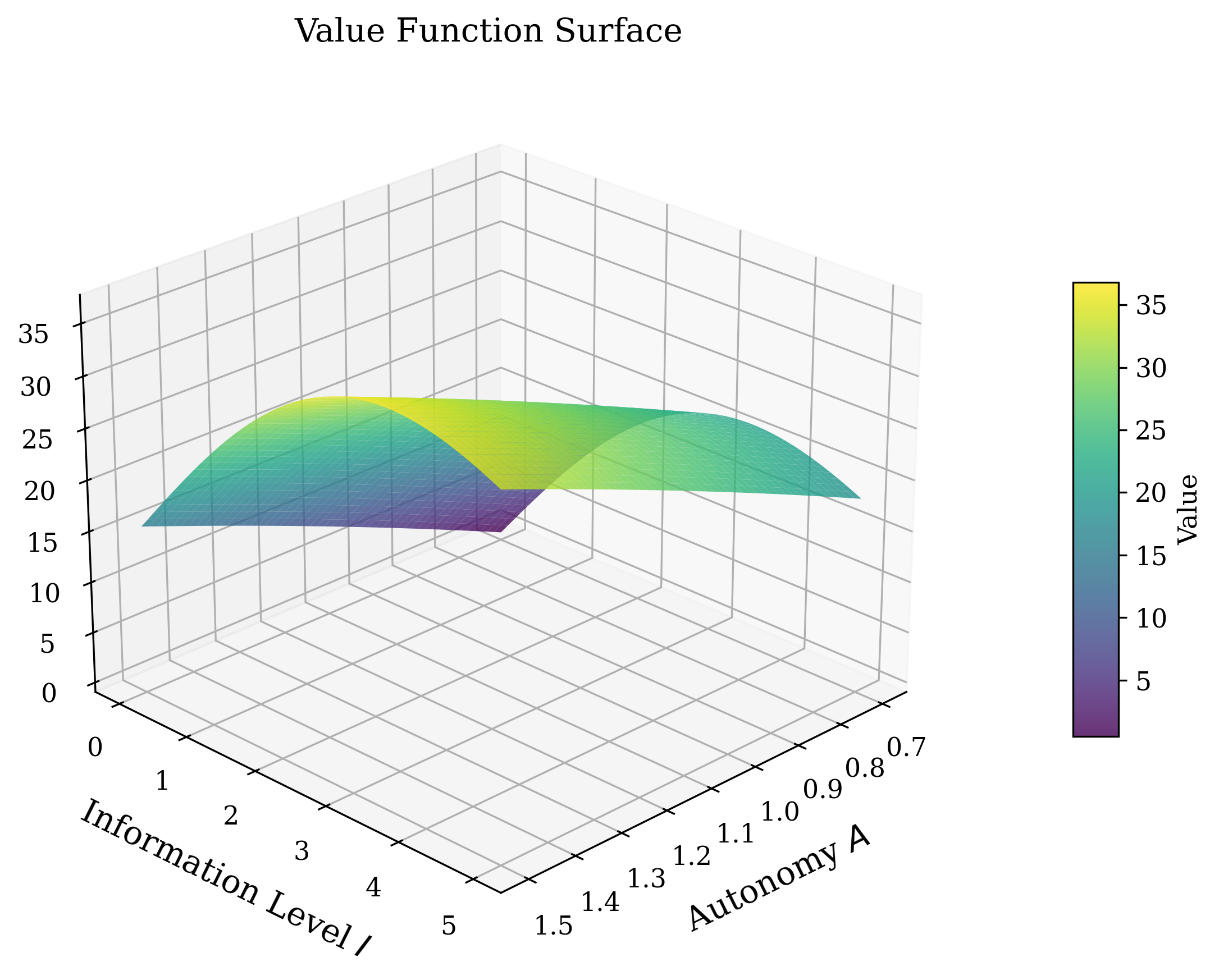}
\caption{Value Function $V(A, I, t=0)$ Exhibits Inverted-U Shape in Information Dimension. The value function peaks at moderate information levels ($I \approx 1.5$), consistent with the transparency paradox. The maximum value occurs at high autonomy with moderate information; the minimum value occurs at boundary autonomy with excessive information.}
\label{fig:value_function}
\end{figure}

\textit{Value Function.} The value function exhibits an inverted-U shape in the information dimension, as shown in Figure~\ref{fig:value_function}.

Maximum value $V(1.5, 1.0, 0) = 46.8$ occurs at high autonomy with moderate information; minimum $V(0.5, 5.0, 0) = -11.2$ occurs at boundary autonomy with excessive information. This formalizes the transparency paradox: moderate transparency optimizes outcomes.

\subsection{Policy Comparison}

\begin{figure}[!htbp]
\centering
\includegraphics[width=0.85\textwidth]{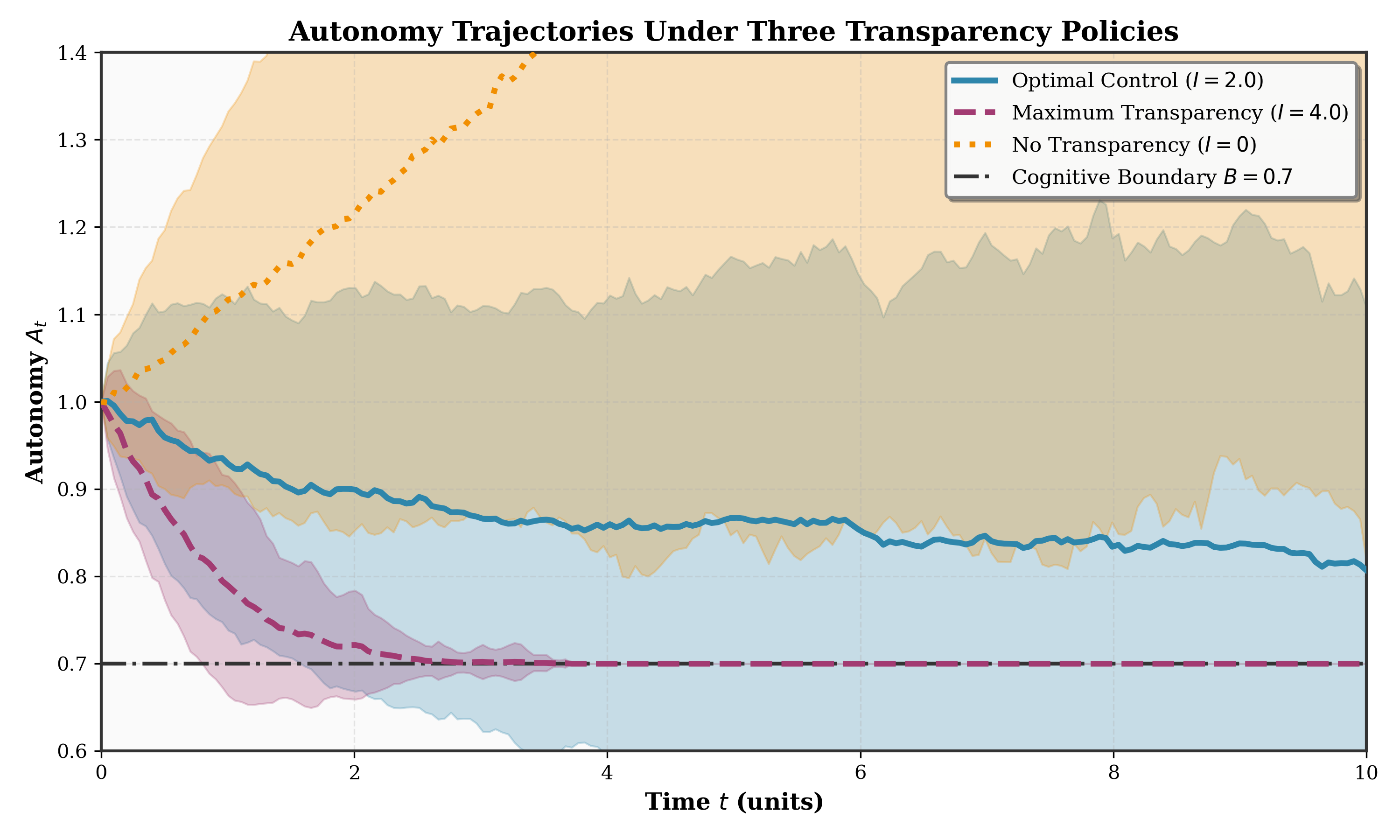}
\caption{Mean Autonomy Trajectories Under Three Transparency Policies. Optimal control (solid line) maintains autonomy while achieving the highest decision quality. Maximum transparency (dashed) depletes autonomy rapidly, leading to high disengagement rates. No transparency (dotted) preserves autonomy but fails to provide decision-relevant information.}
\label{fig:trajectories}
\end{figure}

Monte Carlo simulation (1,000 paths) comparing three policies reveals the advantage of adaptive transparency. Figure~\ref{fig:trajectories} displays mean autonomy trajectories under each policy.

\textit{Optimal Control} ($u_t = u^*(A_t, I_t, t)$): Mean final autonomy 0.99, disengagement probability 47\%, mean quality 7.8.

\textit{Maximum Transparency} ($u_t = u_{\max}$): Mean final autonomy 0.50, disengagement probability 96\%, mean quality 6.4.

\textit{No Transparency} ($u_t = 0$): Mean final autonomy 1.97, disengagement probability 10\%, mean quality 4.1.

Optimal control achieves the highest quality while maintaining autonomy and minimizing disengagement. Maximum transparency improves quality over no transparency, but at a severe autonomy cost. This demonstrates the core tradeoff: aggressive transparency provides information but depletes autonomy; optimal policy navigates this by adapting to cognitive state.

\subsection{Disengagement Timing}

Under maximum transparency ($I = 4.0$), drift is $\mu(4.0) = -0.26$. Proposition 3 predicts $\mathbb{E}[\tau_B] = 2.48$; simulation yields $2.49 \pm 1.12$, confirming mathematical consistency. Scaling to real time (1 unit $\approx$ 10 minutes) predicts disengagement around 25 minutes under high transparency.

\subsection{Working Memory Effects}

Varying WM $\in \{2, 3, 4, 5, 6\}$ demonstrates individual differences in transparency tolerance. Table~\ref{tab:wm_effects} presents the results.

\begin{table}[htbp]
\centering
\caption{Working Memory Effects on Disengagement Under Maximum Transparency}
\label{tab:wm_effects}
\begin{threeparttable}
\begin{tabular}{cccc}
\toprule
Working Memory Capacity & Boundary ($B$) & Expected Time $\mathbb{E}[\tau_B]$ & $P(\tau_B < 10)$ \\
\midrule
2 & 0.70 & 1.29 & .99 \\
3 & 0.60 & 1.84 & .97 \\
4 & 0.50 & 2.49 & .89 \\
5 & 0.40 & 3.30 & .74 \\
6 & 0.30 & 4.33 & .54 \\
\bottomrule
\end{tabular}
\begin{tablenotes}
\small
\item \textit{Note.} Expected disengagement time increases with working memory capacity. Higher capacity yields a lower disengagement boundary, reflecting greater cognitive buffer. Each additional WM unit adds approximately 0.76 time units to expected engagement duration.
\end{tablenotes}
\end{threeparttable}
\end{table}

These results confirm Prediction P4: higher working memory capacity enables more extended engagement under high transparency conditions.

\subsection{Information Level Effects}

Varying constant information levels reveals the inverted-U relationship. Table~\ref{tab:info_effects} presents autonomy and quality outcomes across transparency levels.

\begin{table}[htbp]
\centering
\caption{Information Level Effects on Autonomy and Decision Quality}
\label{tab:info_effects}
\begin{threeparttable}
\begin{tabular}{cccc}
\toprule
Information Level ($I$) & Drift $\mu(I)$ & Expected Autonomy $\mathbb{E}[A_{10}]$ & Decision Quality \\
\midrule
0 & 0.100 & 1.27 & 0.0 \\
1 & 0.040 & 1.09 & 6.7 \\
2 & $-0.040$ & 0.88 & 8.4 \\
3 & $-0.140$ & 0.67 & 7.2 \\
4 & $-0.260$ & 0.45 & 4.8 \\
5 & $-0.400$ & 0.28 & 2.3 \\
\bottomrule
\end{tabular}
\begin{tablenotes}
\small
\item \textit{Note.} Decision quality peaks at $I = 2.0$ despite negative drift, demonstrating the inverted-U relationship between transparency and performance.
\end{tablenotes}
\end{threeparttable}
\end{table}

Quality peaks at $I = 2.0$ despite negative drift, demonstrating an inverted-U shape. Beyond this optimal point, cognitive overload dominates, reducing both autonomy and effective decision quality. This captures the transparency paradox quantitatively: moderate transparency optimizes outcomes, while insufficient or excessive transparency reduces performance.

\subsection{Quantitative Predictions for Empirical Testing}

\textbf{Prediction 1:} Under high transparency ($I > 2.5$), autonomy follows $\mathbb{E}[A_t] = \exp(-0.14t)$.

\textbf{Prediction 2:} Mean disengagement under maximum transparency: $\mathbb{E}[\tau_B] \approx 2.49$ units ($\approx$25 minutes).

\textbf{Prediction 3:} WM effect: $d\mathbb{E}[\tau_B]/d\text{WM} \approx 0.76$ units per WM item.

\textbf{Prediction 4:} Quality-maximizing transparency: $I_{\text{opt}} \approx 2.0 \pm 0.3$.

\textbf{Prediction 5:} Log-autonomy linear in time: $\log A_t - \log A_0 = [\mu(I) - 0.02]t + \text{noise}$.

\section{Discussion}

\subsection{Theoretical Contributions}

This work provides a framework for understanding the transparency paradox by formalizing how cognitive load dynamics determine when AI explanations help or impair. Three contributions advance psychological understanding.

First, a \textit{psychological mechanism explaining the paradox}. By modeling autonomy as a stochastic process influenced by information-induced cognitive load, the framework explains why identical interventions produce opposite effects. The mechanism---transparency triggers metacognitive processing that depletes working memory, reducing control when load exceeds capacity---addresses the contradiction between theoretical advocacy for transparency and evidence of harm. The paradox emerges naturally from cognitive dynamics: excessive information depletes resources faster than it provides value, reducing autonomy until disengagement.

Second, \textit{integration of disparate findings}. The framework explains inverted-U relationships through nonlinear drift with interior optima, timing effects through time-dependent boundaries, and individual differences through capacity parameters. These reflect different state-space regions rather than different mechanisms.

Third, \textit{quantitative predictions enabling falsification}. The five predictions about trajectories, timing, individual differences, and optimal levels provide clear empirical tests. Violations would indicate the need for model refinements (e.g., mean-reverting dynamics, alternative drift functions).

\subsection{Implications for Psychological Theory}

The framework extends cognitive load theory by modeling temporal dynamics of resource depletion, explaining why identical information helps or impairs at different times. It extends autonomy theory by specifying mechanisms through which information affects control and formalizing dynamic depletion. It contributes to dynamic decision-making by showing that rational information acquisition depends on current autonomy and accumulated load, not just information value.

\subsection{Modeling Assumptions and Limitations}

The framework makes simplifying assumptions warranting empirical investigation. The GBM specification captures key phenomena, but alternatives (Ornstein-Uhlenbeck mean-reversion, jump-diffusion, state-dependent volatility) may better fit some data. The quadratic drift captures inverted-U relationships, but other functional forms are possible. The single autonomy dimension ignores potential interactions with trust and engagement. Parameters are theoretically motivated but require empirical estimation.

These assumptions provide clear hypotheses for testing. Model comparison using likelihood ratios and cross-validation will determine which specifications best capture human data. The framework's value lies in providing mathematical language for formalizing transparency effects and generating testable predictions.

The model assumes continuous time dynamics, whereas real human-AI interactions often occur in discrete steps (e.g., per-decision explanations). Future extensions could incorporate discrete-event structures or hybrid continuous-discrete models to better reflect practical scenarios.

\subsection{Practical Implications}

The optimal control solutions yield design principles for adaptive AI systems:

\textit{Dynamic adjustment}: Provide information based on real-time cognitive state. When autonomy is high and information is low, provide detailed explanations; when autonomy declines, reduce transparency.

\textit{Information budgets}: The threshold $I^*(A, t)$ guides how much information users can process. Track cumulative complexity and provide transparency only when approaching limits.

\textit{Personalized thresholds}: Assess cognitive capacity and adjust accordingly. Higher-capacity users tolerate more information; lower-capacity users require conservative policies.

\textit{Temporal adaptation}: Provide progressively more information near deadlines, accepting some autonomy cost because time constraints increase the value of information.

\subsection{Limitations and Future Directions}

The framework requires empirical validation. Laboratory experiments should test the five predictions through repeated autonomy measures, disengagement timing, and WM correlations. Field studies in healthcare, transportation, and finance will assess generalization. Parameter estimation through maximum likelihood fitting will refine quantitative predictions. Empirical validation of predictions P1--P4 is currently underway using a within-subjects design that manipulates explanation complexity while measuring autonomy trajectories and working memory capacity.

\textit{Generalizability beyond AI.} While developed for XAI contexts, the autonomy depletion mechanism applies broadly to information-rich environments. Educational settings exhibit similar dynamics: excessive instructional scaffolding can undermine student autonomy and self-regulation \citep{sweller1988}. Financial decision-making under information overload, medical informed consent processes, and even social media feed algorithms present analogous transparency-autonomy tradeoffs. The HIAG framework's core insight---that information provision triggers metacognitive processing with cognitive costs---transcends AI-specific applications. In non-Western cultural contexts, where collectivist orientations may prioritize relational autonomy over individual control \citep{markus1991}, the model's parameters (e.g., baseline drift $\mu_0$) could be adjusted to reflect varying autonomy preferences, potentially leading to different optimal thresholds.

\textit{Ethical considerations.} Real-time cognitive state monitoring for adaptive transparency raises privacy and autonomy concerns. Systems that infer user mental states from physiological signals or interaction patterns could enable manipulation or surveillance. Design guidelines should ensure that adaptive transparency serves user welfare rather than engagement optimization. Transparency about the transparency system itself---meta-transparency---may be ethically required. Additionally, individual differences in the boundary parameter $B$ raise fairness questions: should systems provide less information to lower-capacity users, potentially limiting their access to decision-relevant data?

\textit{Methodological directions.} Future research should employ eye-tracking to measure metacognitive processing load (e.g., regression patterns, pupil dilation as cognitive effort proxy), experience sampling methods for real-time autonomy assessment, and neuroimaging to identify neural correlates of autonomy depletion. Computational approaches could extend the model to multi-agent settings where AI transparency affects team dynamics.

Necessary extensions include multivariate models incorporating trust dynamics, learning effects as users develop expertise, and cultural differences in autonomy preferences. Recovery dynamics after depletion and emotional moderation of cognitive effects represent additional directions.

\section{Conclusion}

This work provides a framework for understanding the transparency paradox by formalizing how cognitive load dynamics determine when AI explanations help or impair. The core insight: transparency effects depend on psychological state trajectories rather than static design choices. Information provision triggers metacognitive processing that depletes working memory, reducing control when load exceeds capacity.

The mathematical formalization integrates contradictory findings into a common account, generates falsifiable predictions, and yields actionable design principles. Transparency is not a fixed parameter but a dynamic control problem whose solution balances information provision against cognitive preservation under capacity constraints.

The paradox can be understood: explanations help when resources suffice to process information, enhancing quality while maintaining autonomy. Explanations impair performance when resources are depleted, overwhelming capacity faster than providing value. Optimal transparency adapts to this reality, responding to evolving cognitive state rather than assuming universal benefits. This insight---that transparency effects depend fundamentally on cognitive load dynamics---provides a foundation for understanding and optimizing information provision in human-AI interaction.

\section*{Key Points}

\begin{itemize}
\item AI explanations can impair decision-making when cognitive load exceeds working memory capacity, explaining the transparency paradox.
\item Autonomy evolves as a stochastic process with information-dependent drift; the framework generates five testable quantitative predictions.
\item Optimal transparency follows a threshold policy: provide information when autonomy is high, withhold when cognitive resources are depleted.
\item Working memory capacity moderates transparency tolerance, with each additional WM item adding approximately 0.76 time units to engagement duration.
\item Adaptive AI systems should adjust transparency based on real-time cognitive state rather than following fixed policies.
\end{itemize}

\bibliography{references}

\newpage
\section*{Author Biographies}

\noindent\textbf{Ancuta Margondai} is a PhD candidate in Modeling and Simulation at the University of Central Florida (expected December 2026). She received her MS in Modeling and Simulation from UCF in 2024 and is concurrently pursuing an MS in Mathematical Science (expected 2027). Her research focuses on human-AI interaction, physiologically responsive AI systems, and computational modeling of cognitive processes. She works under Dr. Mustapha Mouloua in the Transportation Research Group Laboratory, where she leads an interdisciplinary research team.

\vspace{1em}

\noindent\textbf{Mustapha Mouloua} is a Professor of Psychology at the University of Central Florida and Director of the Transportation Research Group Laboratory. He received his PhD in Psychology from the Catholic University of America in 1992. His research interests include human-automation interaction, vigilance, attention, and human factors in transportation systems. He has over 30 years of experience in human factors research and has authored numerous publications on human performance in automated systems.

\end{document}